\documentclass[sigconf]{aamas} 
\usepackage{balance}

\usepackage{amsmath,amsfonts,bm}

\newcommand{\newterm}[1]{{\bf #1}}

\def\eqref#1{equation~\ref{#1}}

\def\1{\bm{1}}

\DeclareMathAlphabet{\mathsfit}{\encodingdefault}{\sfdefault}{m}{sl}
\SetMathAlphabet{\mathsfit}{bold}{\encodingdefault}{\sfdefault}{bx}{n}

\newcommand{\R}{\mathbb{R}}

\usepackage{hyperref}
\usepackage{graphicx}
\usepackage{url}
\usepackage{booktabs}
\usepackage{caption}
\usepackage{xspace}
\usepackage{dsfont}
\usepackage{wrapfig}
\usepackage{color}
\usepackage{multirow}

\def\eg{{\it e.g.}}

\def\ie{{\it i.e.}}

\renewcommand\footnotetextcopyrightpermission[1]{}

\setcopyright{ifaamas}
\acmConference[AAMAS '23]{Proc.\@ of the 22nd International Conference
on Autonomous Agents and Multiagent Systems (AAMAS 2023)}{May 29 -- June 2, 2023}
{London, United Kingdom}{A.~Ricci, W.~Yeoh, N.~Agmon, B.~An (eds.)}
\copyrightyear{2023}
\acmYear{2023}
\acmDOI{}
\acmPrice{}
\acmISBN{}

\acmSubmissionID{148}

\title{Counterfactual Fairness Filter for\\Fair-Delay Multi-Robot Navigation}

\author{Hikaru Asano$^\ast$}
\thanks{$^\ast$Work done during an internship at OMRON SINIC X}
\affiliation{
  \institution{The University of Tokyo}
  \city{Tokyo}
  \country{Japan}}
\email{asano-hikaru19@g.ecc.u-tokyo.ac.jp}

\author{Ryo Yonetani}
\affiliation{
  \institution{OMRON SINIC X}
  \city{Tokyo}
  \country{Japan}}
\email{ryo.yonetani@sinicx.com}

\author{Mai Nishimura}
\affiliation{
  \institution{OMRON SINIC X}
  \city{Tokyo}
  \country{Japan}}
\email{mai.nishimura@sinicx.com}

\author{Tadashi Kozuno}
\affiliation{
  \institution{OMRON SINIC X}
  \city{Tokyo}
  \country{Japan}}
\email{tadashi.kozuno@sinicx.com}

\newcommand{\methodfull}{Navigation with Counterfactual Fairness Filter\xspace}
\newcommand{\methodabb}{NCF2\xspace}
\newcommand{\problem}{fair-delay multi-robot navigation\xspace}

\renewcommand{\paragraph}[1]{\smallskip\noindent\textbf{#1:}\enskip}

\begin{abstract}
Multi-robot navigation is the task of finding trajectories for a team of robotic agents to reach their destinations as quickly as possible without collisions. In this work, we introduce a new problem: fair-delay multi-robot navigation, which aims not only to enable such efficient, safe travels but also to equalize the travel delays among agents in terms of actual trajectories as compared to the best possible trajectories. The learning of a navigation policy to achieve this objective requires resolving a nontrivial credit assignment problem with robotic agents having continuous action spaces. 
Hence, we developed a new algorithm called Navigation with Counterfactual Fairness Filter (NCF2). With NCF2, each agent performs counterfactual inference on whether it can advance toward its goal or should stay still to let other agents go. Doing so allows us to effectively address the aforementioned credit assignment problem and improve fairness regarding travel delays while maintaining high efficiency and safety. 
Our extensive experimental results in several challenging multi-robot navigation environments demonstrate the greater effectiveness of NCF2 as compared to state-of-the-art fairness-aware multi-agent reinforcement learning methods. Our demo
videos and code are available on the project webpage: \url{https://omron-sinicx.github.io/ncf2/}.
\end{abstract}

         
\newcommand{\BibTeX}{\rm B\kern-.05em{\sc i\kern-.025em b}\kern-.08em\TeX}

\begin{document}

\pagestyle{fancy}
\fancyhead{}

\maketitle 

\section{Introduction}

Time is equally precious to everyone. Having only certain people experience delays in services where efficiency is critical would lead to unfairness. Familiar examples include food delivery and cab dispatch services: imagine a situation in which your afternoon meeting is about to start, but only the food that you ordered has not been delivered. Fairness regarding time is also be important in life-threatening situations, such as disaster relief and evacuation.

We are interested in achieving such temporal fairness among multiple autonomous mobile robotic agents. In particular, this work proposes \emph{\problem}, in which each robot must travel to its destination as fairly as possible in terms of temporal delays, \eg, by avoiding a situation where one agent is an hour late in arriving while the other agents are only 10 minutes late. In the fields of AI and robotics, multi-robot navigation has long been studied as a practical application of multi-agent reinforcement learning (MARL)~(\eg, \cite{lowe2017multi,wang2020model}) or multi-agent pathfinding (MAPF)~\cite{stern2019multi}. Typically, each robot must navigate as quickly as possible to its destination while avoiding collisions with obstacles (\eg, walls, shelves) and other robots in motion. Furthermore, the layout of obstacles may have changed since an environmental map was first constructed. Even in such unknown environments, the robots still have to cooperate with each other to reach their destinations efficiently and safely.

\begin{figure*}[t]
    \centering
    \includegraphics[width=\linewidth]{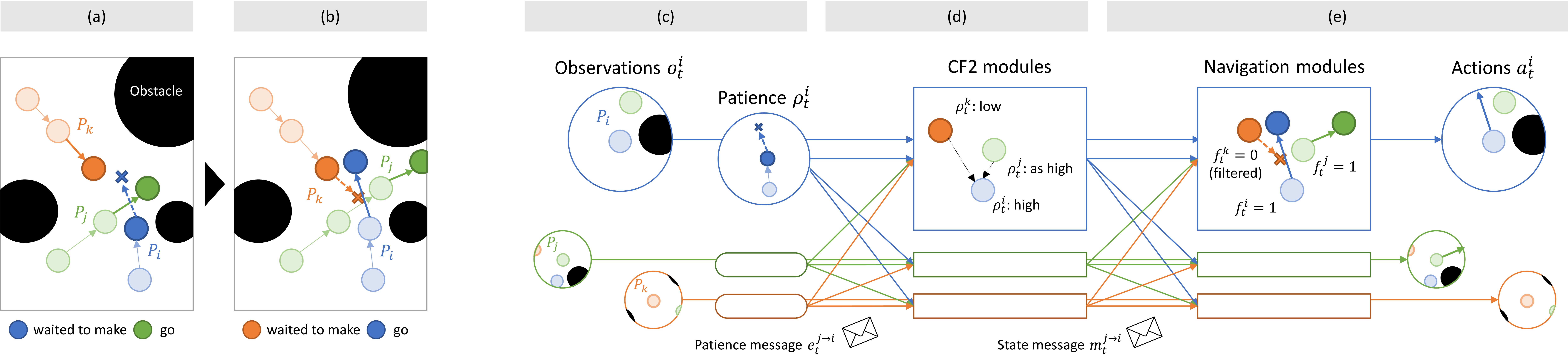}
    \caption{Example of fair-delay multi-robot navigation by \methodabb. (a, b) Three agents perform a navigation task while giving way to equalize their travel delays. (c) To do so, the agents first observe the situations around their current positions and share their ``patience,'' which represents how much delay they have experienced so far. (d) The CF2 module processes the patience information to decide and share whether each agent should move. (e) The agents take their next actions according to the navigation module.}
    \label{fig:teaser}
\end{figure*}

\looseness=-1
By considering fairness for travel delays on top of such a navigation objective, \problem can be viewed as a novel extension of fairness-aware MARL problems for robot navigation tasks. Specifically, we aim to \emph{find navigation policies that can equalize the delays among agents between their actual trajectories and the best possible trajectories they could have taken by ignoring the presence of other agents, while also pursuing high efficiency and safety.}

Achievement of the above goal requires solving a challenging credit assignment problem of evaluating the impact of an agent's actions on the success of other agents with continuous action spaces~\cite{Yang2020CM3}. In multi-robot navigation, situations that require fairness consideration emerge naturally but unpredictably. An agent giving way in one place may cause another agent's travel to become blocked at another time in another place. For prior fairness-aware MARL methods that simply learn fairness through a reward~\cite{jiang2019learning,zimmer2021learning}, it would be hard to reason about which actions contribute to fairness. Such reasoning could be possible via \emph{counterfactual inference}, which compares actual events to events that did not occur but could have. Nevertheless, existing counterfactual methods for multi-agent tasks often assume only discrete action spaces~\cite{li2021shapley,yuan2022counterfactual}, which makes it nontrivial how to define \emph{default actions} (\ie, replacement actions for when agents do not take planned actions) for robotic agents that are typically commanded with continuous values. Furthermore, most counterfactual methods require centralized training with a global critic evaluating the actions of all agents~\cite{foerster2018counterfactual,su2020counterfactual,li2021shapley}, which is not feasible as the number of agents increases.

\looseness=-1
To address this challenge, we propose a decentralized, counterfactual, multi-robot navigation algorithm called \newterm{\methodfull (\methodabb)}. The key idea is to equip each agent with a policy comprising a \emph{navigation module} and a \emph{counterfactual fairness filter (CF2) module}. While the navigation module enables agents to take continuous actions to reach their destinations efficiently and safely, the CF2 module enables them to perform counterfactual inference on action choices to improve fairness. Fig.~\ref{fig:teaser} illustrates a typical case targeted by \methodabb. In Fig.~\ref{fig:teaser}b, agent $P_i$ wants to advance according to its navigation module, while agent $P_k$ is trying to cut in along the way. As shown in Fig.~\ref{fig:teaser}a, $P_i$'s travel already involved a delay to make room for another agent $P_j$. Here, each agent's CF2 module observes the local situations around the agent and asks $P_k$ to suspend its next action and let $P_i$ go ahead. By repeatedly resolving such small problems in every local situation, we can expect to eventually achieve fairness across all agents.

\looseness=-1
In \methodabb, agents learn a shared policy in a decentralized fashion by communicating with each other as follows. At each timestep, each agent partially observes the environment around its current position and creates messages about its \emph{`patience'}. This notion of patience indicates how much delay an agent has experienced so far in comparison to the best possible actions it could have taken(Fig.~\ref{fig:teaser}c). An agent's CF2 module receives the patience information and decides whether agents should move according to the difference in the patience scores between each agent and its surroundings (Fig.~\ref{fig:teaser}d). The CF2 module outputs are again shared across agents, and the navigation module determines the actual next action while considering which agents are allowed to move (Fig.~\ref{fig:teaser}e). As a result of the action, each agent receives a reward for efficiency, safety, and the counterfactual measurement of fairness. The experience from this procedure is accumulated over all agents and steps, and it is used to learn the policy to maximize the cumulative rewards via a reinforcement learning (RL) algorithm.

To evaluate the effectiveness of our approach, we constructed a set of challenging, fair-delay, multi-agent navigation environments involving multiple wheeled robotic agents with a simple but practical kinematics model and a continuous action space. Our results show that the policy learned with \methodabb outperforms state-of-the-art fairness-aware MARL methods \citep{jiang2019learning,zimmer2021learning} in terms of the success rate, travel time, and degree of fairness.
\section{Problem Statement}
\label{sec:preliminaries}
This section introduces the proposed \problem problem step by step.
We address a multi-agent setting with a swarm of $N$ robotic agents, $P_1, \ldots, P_N$, to simulate autonomous mobile robots in a 2D environment with some static obstacles.
We begin by explaining a simpler case with only a single agent $P_i$ and then extend it to the \problem problem.

\paragraph{Single-Robot Navigation}
\looseness=-1
This problem is formulated as a tuple $\Pi_i = (\mathcal{S}_i, \mathcal{A}_i, \mathcal{O}_i, I_i, T_i, R_i, \Omega_i, t_{\mathrm{max}})$,
where $\mathcal{S}_i$, $\mathcal{A}_i$, and $\mathcal{O}_i$ are the (individual) state, action, and observation spaces, respectively.
\footnote{We omit ``individual'' when it is clear from context.}
Specifically, $\mathcal{S}_i = \textrm{SE(2)} = \mathbb{R}^2 \times \mathbb{S}^1$, where $\textrm{SE(2)}$ is the special Euclidean group whose elements represent the robot's position and orientation.
It comprises a set of invalid states, $\mathcal{I}_i$, and a set of valid states, $\mathcal{V}_i$.
Invalid states are those where the agent is too close to an obstacle and is considered to be crashing into it, whereas all other states are valid.
\footnote{The definition of a ``crash'' depends on the environment. All environment-specific details, such as the definitions of a crash and the goal region, are given in Sec.~\ref{sec:env}.}
At the beginning of an ``episode,''  a start state $s_1^i$ and a goal region center $c_i \in \mathbb{R}^2$ are sampled from a probability distribution $I_i$ over $\mathcal{V}_i \times \mathbb{R}^2$.
A ball (whose radius is environment-dependent) around $c_i$ is defined as the goal region, and valid states whose position belongs to the goal region are called goal states.
The set of all goal states is denoted by $g_i \subset \mathcal{V}_i$.
The agent's aim is to reach a goal state without entering $\mathcal{I}_i$.
At each timestep $t$, the agent receives partial information $o_t^i = \Omega_i (s_t^i) \in \mathcal{O}_i$ about the environment around its current location, where $s_t^i$ is the current state.
Then, it chooses and executes an action $a_t^i \in \mathcal{A}_i \subset \R^2$, which entails the robot's velocity and angular velocity.
As a consequence, the state $s_t^i$ transitions to $s_{t+1}^i = T_i (s_t^i, a_t^i, c_i)$, and the agent receives a reward $r_t^i = R_i (s_t^i, a_t^i, s_{t+1}^i, c_i)$.
When the agent reaches an invalid or goal state, its next state is the same regardless of any action, \ie, $T_i (s_t^i, a_t^i, c_i) = s_t^i$.
Each episode is reset at $t+1 = t_{\mathrm{max}}$.
If a trajectory, $\tau_i=(s^i_t)_{t=1}^{t_{\mathrm{max}}}$, ends at a goal state, the trajectory is considered valid.
The length of a valid trajectory is defined as $l_i=\min \{ t | s^i_t \in g_i \}$, which indicates the travel time.
For simplicity, we assume a \emph{homogeneous} setup where $\Pi_1 = \cdots \Pi_N = \Pi'$.
Because this formulation can be understood as a goal-conditional version~\citep{andrychowicz2017hindsight} of a partially observable Markov decision process (POMDP)~\citep{kaelbling1998planning}, we refer to it as an individual POMDP (for $P_i$).

\paragraph{Multi-Robot Navigation}
\looseness=-1
This problem is formulated as a tuple $\Pi = (N, \mathcal{S}, \mathcal{A}, \mathcal{O}, I, T, R, \Omega, t_{\mathrm{max}})$, where $\mathcal{S} = \times_{i=1}^N \mathcal{S}_i$, $\mathcal{A} = \times_{i=1}^N \mathcal{A}_i$, and $\mathcal{O} = \times_{i=1}^N \mathcal{O}_i$ are the (joint) state, action, and observation spaces, respectively.
\footnote{We omit ``joint'' when it is clear from context.}
\footnote{Without loss of generality, we assume that each $\mathcal{O}_i$ contains not only observations in the single-agent case but also the multi-agent case considered here. Indeed, an agent's observation is the same in both cases when there are no neighboring agents.}
While the state space can be partitioned as before into a set of invalid states, $\mathcal{I}$, and a set of valid states, $\mathcal{V}$,
$\mathcal{I}$ now also includes states where an agent is crashing into another agent.
An initial state $s_1 = (s_1^i)_{i=1}^N \in \mathcal{V}$, a goal center $c = (c_i)_{i=1}^N \in (\mathbb{R}^2)^N$, and a goal state $g = (g_i)_{i=1}^N \in \mathcal{V}$ are determined as before.

\footnote{To avoid pathological cases, we assume that the goal centers are sufficiently separated from each other.}
The probability distribution $I$ of $(s_1, c)$ is simply chosen to be consistent with this construction.
The aim of the agents in the swarm is to reach $g$ without entering $\mathcal{I}$.
At each timestep $t$, the agents receive an observation $o_t = (o_t^i)_{i=1}^N = \Omega (s_t) \in \mathcal{O}$, where $s_t = (s_t^i)_{i=1}^N \in \mathcal{S}$ is the current state.
Then, the agents exchange messages, possibly multiple times, to cooperate with each other. Each message from $P_j$ to $P_i$ at timestep $t$ is given by a $D$-dimensional real-valued vector $m^{j\rightarrow i}_t \in \mathbb{R}^D$.
After message exchanges, the agents choose and execute an action $a_t = (a_t^i)_{i=1}^N \in \mathcal{A}$.
As a consequence, the state $s_t$ transitions to $s_{t+1} = (s_{t+1}^i)_{i=1}^N = T (s_t, a_t, c)$, and the agents receive a global reward $r_t = R (s_t, a_t, s_{t+1}, c)$.
We assume that $s_{t+1}^i = T_i (s_t^i, a_t^i)$, unless an agent (say $P_i$) is crashing into another agent, in which case $s_{t+1}^i = s_t^i$.
Each episode is reset at $t+1 = t_{\mathrm{max}}$. 
A trajectory $\tau=(s_t)_{t=1}^{t_{\mathrm{max}}}$ is valid if it ends at a goal state, \ie, if all agents arrived at their own goal.
The length of a valid trajectory is given by $l=\min \{ t | s_t \in g \}=\max \{\min\{t_i | s^i_t \in g_i\}\}=\max \{l_i\}$, where $l_i$ here is the travel time of $P_i$. Note that this trajectory length is referred to as a \emph{makespan} in the context of MAPF~\cite{stern2019multi}.
This formulation can be viewed as a goal-conditional version of a decentralized POMDP \citep{goldman2004decentralized}, hereafter simply called a Dec-POMDP.

\paragraph{Cooperative Policy}
\looseness=-1
Given an observation $o^i_t\in\Omega_i$ and a collection of messages from neighboring agents, $\{ m^{j\rightarrow i}_t\mid j \in \mathcal{N}^i_t \}$, where the $N^i_t\subset\{1,\ldots,N\}$ are the indices of agents regarded as $P_i$'s neighbors at time $t$, each agent executes an action $a_t^i$ sampled from a shared \emph{cooperative policy} $\pi (\cdot \mid o^i_t, c_i, \{m^{j\rightarrow i}_t\mid j \in \mathcal{N}^i_t\})$ to move one step toward the goal $g_i$. If $\pi$ is trained properly and is running for an episode with a start state $s_1 = (s_1^i)_{i=1}^N$ and a goal state $g = (g_i)_{i=1}^N$, then a valid trajectory $\tau^\pi$ with length $l^\pi=\max\{l^\pi_i\}$ is obtained.

\paragraph{Delay Measurement with Solitary Policy}
For a valid trajectory $\tau^\pi$, each agent's goal time $l^\pi_i$ can be delayed in comparison to the case when agents do not have to avoid collisions with each other. To quantify this delay, we assume that another deterministic policy, the \emph{solitary} policy $\mu: \mathcal{O}_i \times c_i \to \mathcal{A}_i$, is pre-trained for an individual POMDP $\Pi_i$ to ensure safe, efficient actions in the absence of other agents. By considering the interaction of a single agent $P_i$ with the multi-robot environment $\Pi$ and the solitary policy $\mu$ while freezing the other agents, we obtain the \emph{the best possible trajectory} $\tau^\mu_i$, with length $l^\mu_i$, that $P_i$ could have taken. (The start and goal states here are $s_1^i$ and $g_i$, respectively.) Then, the delay for agent $P_i$ with the cooperative policy $\pi$ can be calculated as $\delta_i=l^\pi_i-l^\mu_i$.

\paragraph{Fair-Delay Multi-Robot Navigation Problem}
For a problem instance modeled by a Dec-POMDP $\Pi$, the navigation task is regarded as \emph{successful} if we can find a valid cooperative trajectory $\tau^{\pi}$. By contrast, navigation fails if any agent fails to reach its goal within $t_{\mathrm{max}}$ or collides with a static obstacle, or if any two agents collide with each other. For a successful task, we define the degree of (un)fairness in terms of the \emph{the variance of delays}, \ie, $\frac{1}{N}\sum_i (\delta_i - \frac{1}{N}\sum_j \delta_j)^2$: the smaller this variance is, the fairer the solution is. We also measure the travel (in)efficiency in terms of the valid trajectory length $l^\pi$ (\ie, the makespan), where a shorter length is better.

The goal of the proposed \emph{fair-delay multi-robot navigation} is to find a cooperative policy that is successful with the highest probability possible (\ie, the policy achieves a high safety level) while also making the unfairness and inefficiency scores as low as possible for the successful instances.
\section{\methodabb: \methodfull}
We construct the proposed cooperative policy, hereafter referred to as the \emph{\methodabb policy}, via two modules. The \emph{navigation module} produces agent actions in a continuous space for travel efficiency and safety, while the \emph{counterfactual fairness filter (CF2) module} judges whether agents should move by considering fairness and efficiency.
The CF2 module's action is restricted to be binary, which makes it easy to incorporate counterfactual reasoning while allowing agents to move with the continuous action space. Below, we first describe how to generate fairness-aware actions with the \methodabb policy in Sec.~\ref{sec:generating}; we then introduce reward designs to learn the policy in a decentralized fashion via an RL algorithm in Sec.~\ref{sec:learning}.

\subsection{Generation of Fairness-Aware Actions}
\label{sec:generating}
The \methodabb policy generates actions in the following three steps, with multiple message exchanges between agents (see also Fig.~\ref{fig:teaser}).

\paragraph{Step 1: Calculate Patience as Proxy for Delays (Fig.~\ref{fig:teaser}c)}
The first step is to measure the delay $\delta_i=l^\pi_i - l^\mu_i$ as defined in Sec.~\ref{sec:preliminaries}. However, the exact value is unavailable during task execution because the actual trajectory length $l^{\pi}_i$ is unknown until each agent arrives at the goal. Instead, we introduce a proxy measurement, referred to as \emph{patience}, which indicates how patient an agent has been in being cooperative (\ie, moving by communicating with others) rather than solitary (\ie, moving while ignoring others). We calculate the patience as follows by using the action-value function of the solitary policy $\mu$ on an individual POMDP $\Pi_i$, \ie, $Q_\mu$: 
\begin{equation}
   \rho^i_t = \sum_{t'=1}^{t-1} \left(Q_{\mu}(o^i_{t'}, \hat{a}^i_{t'}, c_i) - Q_\mu(o^i_{t'}, a^i_{t'}, c_i)\right),
   \label{eq:patience}
\end{equation}
where $\hat{a}^i_{t'} = \mu (o_{t'}^i, c_i)$ and $a^i_{t'}\sim\pi(\cdot \mid o_{t'}^i, c_i, \{m^{j\rightarrow i}_t\mid j\in \mathcal{N}^i_{t'}\})$ are the actions produced by the solitary and cooperative policies, respectively. Because $\mu$ is trained to enable efficient navigation in a single-agent environment, as assumed in the previous section, its action-value function evaluates how promising input actions are from this perspective. As a small $Q_\mu(o^i_{t'}, a^i_{t'}, c_i)$ indicates that action $a^i_{t'}$ will not move the agent efficiently toward the goal, a large $\rho^i_t$ implies that the delay will be as large.

\paragraph{Step 2: Process Patience with CF2 Module (Fig.~\ref{fig:teaser}d)}
\looseness=-1
The patience is included in a message that is shared among the agents and used in the CF2 module to judge whether each agent should really move. We define this message, hereafter referred to as a \emph{patience message}, as follows:
\begin{equation}
    e^{j\rightarrow i}_t
    =
    \left(
        \Delta(s^j_t, s^i_t);
        \frac{\rho^j_t - \rho^i_t}{\sum_{j\in\mathcal{N}^i_j} \rho^j_t}; \Delta(\hat{s}^j_{t+1}, s^i_t); \frac{\rho^j_t - \rho^i_t}{\sum_{j\in\mathcal{N}^i_j} \rho^j_t}
    \right).
    \label{eq:extended}
\end{equation}
Here, $\Delta(a, b); a, b \in \mathcal{S}_i$ is the relative position of state $a$ with respect to $b$, $\hat{s}^i_{t+1}$ is the next state sampled from an individual (\ie, single-agent) POMDP $\Pi'$ with action $\hat{a}^i_t$, and $(u; v)$ is a vector obtained by vertically concatenating column vectors $u$ and $v$. Intuitively, a patience message expresses where each agent is and where it would go if it were allowed to ignore other agents, given the current relative patience. We model the CF2 module as a sub-policy $\psi(\cdot \mid o^i_t, c_i, \{e^{j\rightarrow i}_t\mid j\in\mathcal{N}^i_t\})$, which provides a binary decision $f^i_t\in \{0, 1\}$ based on the current observation and patience messages. Once learned properly, it gives $f^i_t = 1$ to allow agents that have been sufficiently patient to move, and it acts like a message filter in the next step. 

\paragraph{Step 3: Decide Next Actions (Fig.~\ref{fig:teaser}e)}
\looseness=-1
Given the output $f^i_t$ from the CF2 module, we define a \emph{state message} as follows:
\begin{equation}
m^{j\rightarrow i}_t=\left(f^j_t\Delta(s^j_t, s^i_t); f^j_t\Delta(\hat{s}^j_{t+1}, s^i_t)\right),
\label{eq:filtered}
\end{equation}
which represents the relative state of agent $P_j$ with respect to $P_i$ if $P_j$ is allowed to move (\ie, $f^j_t = 1$) or contains all-zeros otherwise. This message is again exchanged among the agents, and used to generate each agent's the next action. The navigation module $\phi$ is modeled as another sub-policy to sample a continuous action $a^i_t$ from an observation $o^i_t$ and state messages $\{m^{j\rightarrow i}_t\mid j \in \mathcal{N}^i_t\}$. The sampling is filtered by $f^i_t$ to stop agents that are not allowed to move:
\begin{equation}
   a^i_t = f^i_t \cdot \nu,\;\; \nu \sim \phi\left(\cdot \mid o^i_t, c_i, \left\{m^{j\rightarrow i}_t\mid j \in \mathcal{N}^i_t\right\}\right).
   \label{eq:next_action}
\end{equation}

\paragraph{Encoding Messages} In the above action generation procedure, collections of messages $\{e^{j\rightarrow i}_t\mid j \in \mathcal{N}^i_t\}$ and $\{m^{j\rightarrow i}_t\mid j \in \mathcal{N}^i_t\}$ are each encoded into a single fixed-length vector representation via the scaled dot-product attention mechanism~\cite{vaswani2017attention} (see Appendix B.1 for more details). This is an effective approach for handling sets with variable numbers of messages in multi-agent modeling~\cite{das2019tarmac}.

\subsection{Decentralized Learning of \methodabb Policy}
\label{sec:learning}
The \methodabb policy is learned in a decentralized fashion to ensure scalability for the number of agents. Specifically, to improve the fairness regarding delays while maintaining navigation efficiency and safety, we design a local reward $r^i_t$ that is given to each agent, instead of explicitly designing a global reward function $R$ for the Dec-POMDP $\Pi$.

Crucially, the navigation and CF2 modules each act as sub-policies that pursue different objectives, and thus, they should be evaluated differently. To this end, we use the hybrid reward architecture~\citep{van2017hybrid} and define the local reward $r^i_t$ as the sum of two rewards: an \emph{efficiency-safety reward} and a \emph{fairness-efficiency reward}. As explained below, they are given \emph{separately} to the two modules. 

\paragraph{Efficiency-Safety Reward}
The navigation module is trained with the following reward:
\begin{equation}
\hat{r}^i_t= 
\begin{cases}
r_\mathrm{goal} - r_\mathrm{time} & \textrm{upon reaching the goal} \\ 
- r_\mathrm{crash} - r_\mathrm{time} & \textrm{if crashing into an obstacle or another agent} \\ 
0 & \textrm{after reaching the goal or crashing}\\
 - r_\mathrm{time} & \textrm{otherwise.}
\end{cases}
\label{eq:safe-efficient-reward}
\end{equation}
Here, $r_\mathrm{goal}>0$ and $r_\mathrm{crash}>0$ are respectively the goal reward and crash penalty given sparsely, and $r_\mathrm{time}>0$ is a time penalty to encourage agents to reach the goal in the shortest possible time.

\paragraph{Fairness-Efficiency Reward}
On the other hand, the CF2 module is given a counterfactual reward to improve fairness while maintaining efficiency. We consider the most cooperative action that could happen if all agents were allowed to move by their CF2 modules as a \emph{default action}, \ie, a replacement for the planned action:
\begin{equation}
   \bar{a}^i_t \sim \phi\left(\cdot \mid o^i_t, c_i, \left\{\left(\Delta(s^j_t, s^i_t), \Delta(\hat{s}^j_{t+1},  s^i_t)\right)\mid j\in \mathcal{N}^i_t\right\}\right).
\end{equation}
This enables us to define the \emph{improvement} for an actual action $a^i_t$ by applying the action-value function of the solitary policy $Q_\mu$ to an individual POMDP $\Pi_i$ as follows:
\begin{equation}
   \xi^i_t =  Q_\mu(o^i_t, a^i_t, c_i) - Q_\mu(o^i_t,  \bar{a}^i_t, c_i).
\end{equation}
Similarly to how the patience $\rho^i_t$ is characterized by $Q_\mu$ in Eq.~(\ref{eq:patience}), this improvement $\xi^i_t$ indicates how much better the actual action $a^i_t$ has become compared to the default action, by preventing some agents from moving via $m^{j\rightarrow i}_t$ in Eq.~(\ref{eq:next_action}). It is then aggregated across neighboring agents, while being weighted by the relative patience, $\rho^j_t - \rho^i_t$, to form a sparse \emph{fairness-efficiency} reward that is given when a CF2 module takes $f^i_t=0$:
\begin{equation}
    \tilde{r}^i_t= (1-f^i_t) \left(\alpha \frac{\sum_{j\in\mathcal{N}^i_t}(\rho^j_t - \rho^i_t)\cdot \xi^j_t}{\sum_{j\in\mathcal{N}^i_j} \rho^j_t} - \beta \frac{\rho^i_t}{\sum_{j\in\mathcal{N}^i_j} \rho^j_t}\right).
    \label{eq:fair-efficient-reward}
\end{equation}
Intuitively, this reward becomes higher for agents that decide to stop when their surrounding agents have been more patient and also can improve their action values by moving. The second term is a penalty to prevent excessive encouragement of agents to stop and sacrifice efficiency. Here, $\alpha,\beta$ are user-specified constants to control the balance between fairness and efficiency. Further analysis of this reward design is presented in Appendix A.

\section{Experiments}
\subsection{Environments}
\label{sec:env}

Our work targets multi-robot navigation tasks in which robotic agents with practical kinematics move in a 2D environment with some static obstacles. This kind of task requires a new multi-agent environment that is completely different from previous ones that only support discrete movements on a grid map, as what was used in \cite{jiang2019learning}. To this end, we extend a multi-agent path planning environment~\cite{okumura2022ctrms,jaxmapp_2022} with crowd-aware single-robot navigation~\cite{chen2019crowd} to construct environments with the following features.

\paragraph{Agent Design} Each agent simulates a wheeled robot equipped with a simulated lidar sensor. The sensor perceives the distances to surrounding obstacles and other agents at equal-angle intervals with a maximum scan range of $0.1\times\texttt{map-size}$ to obtain 64-dimensional scan data. The robot is modeled as a circle of fixed size ($0.02\times\texttt{map-size}$), and we determine that a collision occurs if an obstacle or another agent enters the circle. The agent can exchange messages with other agents within a range of $0.15\times\texttt{map-size}$. Finally, an agent's motion follows a non-holonomic kinematics model that is characterized by a maximum velocity ($0.05\times \texttt{map-size/timestep}$) and a maximum angular velocity ($0.25\pi\texttt{/timestep}$). The agent accepts continuous-valued velocity and angular velocity commands to move forward.

\begin{figure*}[t]
    \centering
    \includegraphics[width=.9\linewidth]{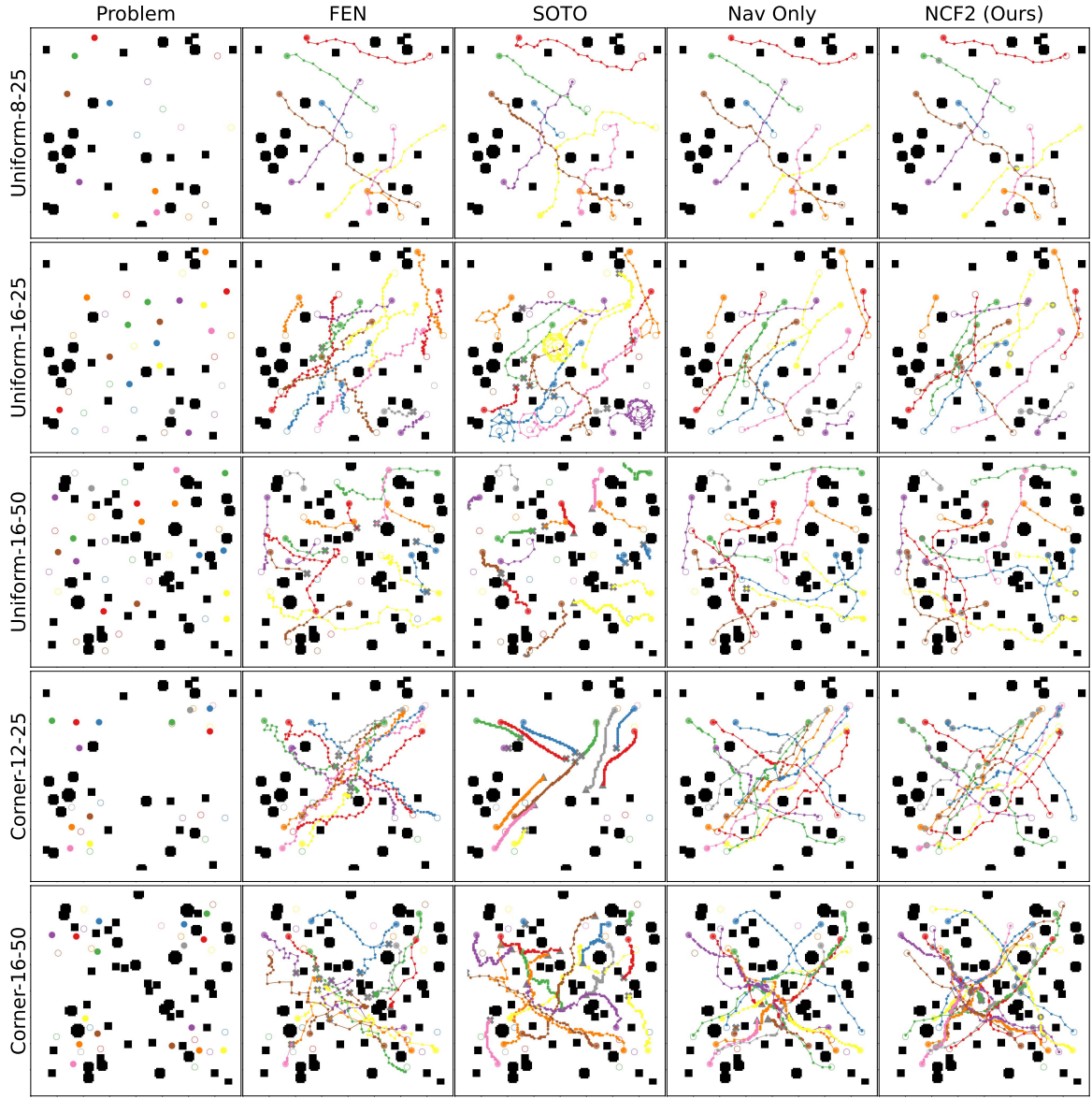}
    \caption{Qualitative comparisons of the methods and environments (best viewed in color). Start positions and goal centers are represented by filled and white circles, respectively, with different colors for different agents. The histories of the agent positions are indicated by small circles connected by line segments. For the case of \methodabb, moments when the CF2 module decided to stop an agent are annotated by gray circles. Collisions and timeouts are shown by gray crosses and triangles respectively. The black circles are obstacles.}
    \label{fig:results}
\end{figure*}

\paragraph{Environment Design} The environment is a map of size $128\times 128$ (\ie, $\texttt{map-size}=128$) and includes static circular obstacles. In our experiments, the goal radius was set to $0.02\times \texttt{map-size}$, and the time limit was set to $t_\mathrm{max}=100$. As shown in Fig.~\ref{fig:results}, we constructed the following four environments:
\begin{itemize}
    \item \textbf{\textsc{Uniform-$N$-25}} comprised $N\in\{8, 12, 16\}$ agents with start and goal positions sampled uniformly in the environment. It included 25 circular obstacles of random size ($[0.05, 0.08]\times \texttt{map-size}$) and location.
    \item \textbf{\textsc{Corner-$N$-25}} also comprised $N$ agents and 25 random circular obstacles, but the agent's start positions were sampled randomly from the four corner regions, and the goal positions were diagonally opposite the starts. This environment requires agents to give way to each other in the center, thus giving a more challenging case than \textsc{Uniform-$N$-25}.
    \item \textbf{\textsc{Uniform-$N$-50}} and \textbf{{Corner-$N$-50}} increased the number of random obstacles from 25 to 50, making the navigation task more difficult.
\end{itemize}

\subsection{Implementation Overview}
Because of space limitations, we give only a brief overview of the implementation here. Further details are described in Appendix B.

\paragraph{Navigation Module}
\looseness=-1
We constructed the navigation module by extending a state-of-the-art RL-based navigation algorithm called residual reactive navigation~\citep{rana2020residual}, which is an application of residual RL~\citep{silver2018residual,johhnnink2018residual} to navigation tasks. Specifically, we used the classical dynamic window approach (DWA)~\citep{fox1997dynamic} as a base controller, and we then applied learning of a residual policy given by a multi-layer perceptron (MLP), which added residual actions to DWA's outputs. An observation of this module, $o^i_t$, comprised the current state, a lidar scan as described in Sec.~\ref{sec:env}, the relative displacement of the goal position with respect to the current position, and the DWA output.

\paragraph{CF2 Module} The CF2 module was also modeled by a standard MLP but had a binary output given by the softmax activation. An observation was the same as for the navigation module.

\paragraph{Solitary Policy} The solitary policy had the same form as that of the navigation module, but without taking messages from other agents as input.

\paragraph{Reward} The reward and penalty values for the safety-efficiency reward in Eq.~(\ref{eq:safe-efficient-reward}) were $r_\mathrm{goal}=3.0, r_\mathrm{crash}=10.0$, and $r_\mathrm{time}=0.1$, such that crashes were penalized more than timeouts. The constants $\alpha, \beta$ in the fairness-efficiency reward in Eq.~(\ref{eq:fair-efficient-reward}) were set empirically to $\alpha=0.5, \beta=0.1$, where $\alpha$ was selected from $\alpha\in [0.1, 5]$ in our pilot experiment.

\paragraph{Learning Algorithm} We used soft actor-critic (SAC)~\citep{haarnoja2018soft, haarnoja2018softapplication} as the RL algorithm to achieve high sample efficiency and learning stability. Crucially, our implementation adopted the distributed experience-generating strategy of Ape-X~\citep{horgan2018distributed} to achieve high data throughput, which we found effective for alleviating the non-stationarity of gathered experiences in off-policy MARL. To learn the \methodabb policy, we first warmed up the navigation module by itself for 1M iterations and then conducted learning by both modules for 1M iterations. This allowed the CF2 module to make a valid decision to pursue fairness by leveraging the safety and efficiency provided by the trained navigation module. The solitary policy was trained for 1M iterations by limiting the number of agents to one in the learning environment. All experiments were conducted with a single GPU (NVIDIA V100) and a single CPU (Intel Xeon Gold 6252).

\subsection{Baselines}
We used the following state-of-the-art fairness-aware MARL methods as baselines.
\begin{itemize}
   \item \textbf{Fair-Efficient Network (FEN)~\cite{jiang2019learning}} is a hierarchical MARL algorithm that simultaneously learns fairness and efficiency. FEN comprises a controller and multiple sub-polices, where the controller first selects a sub-policy and then the selected sub-policy outputs an action. While one of the sub-policies is learned with an efficiency-safety reward $\hat{r}^i_t$ as defined in Eq.~(\ref{eq:safe-efficient-reward}), the remaining sub-policies receive the log-probability of being selected by the controller as a reward. Furthermore, the average reward over the elapsed timesteps is calculated as a utility, and to pursue fairness, the controller is learned to minimize the difference between its own utility and the average utilities of all agents.
   
   \item \textbf{Self-Oriented Team-Oriented Network (SOTO)~\cite{zimmer2021learning}} is another hierarchical algorithm, which seeks to maximize the generalized Gini social welfare function (GGF) to simultaneously quantify efficiency and fairness. SOTO learns two policies: a self-oriented policy and a team-oriented policy. The self-oriented policy first samples an action, which is then used as an input to the team-oriented policy to sample a cooperative action. While the self-oriented policy is learned from the efficiency-safety reward, the team-oriented policy receives as a reward comprising the weighted sum of the other agent's cumulative rewards to maximize the GGF. Here, the weights are determined such that agents with larger cumulative rewards are given smaller values.
\end{itemize}
Note that in our initial study, we realized that direct adaptations of the above methods did not solve our navigation tasks at all. To mitigate this problem, we extended these methods by (1) replacing the base RL algorithm from proximal policy optimization (PPO)~\citep{schulman2017proximal} with our SAC implementation to achieve high data throughput; (2) limiting each agent's communication to a few agents in its vicinity; and (3) using the warmed-up navigation module as one of these methods' sub-policies. For another baseline, we used a degraded version of our approach, referred to as \textbf{Nav-Only}, which used only the trained navigation module to control the agents. To make the experiment fair, the hyperparameters of the RL algorithm were identical to those of the proposed method. The number of training iterations was doubled for Nav-Only to compensate for the warm-ups of the other methods.

\subsection{Evaluation Metrics}
\label{sec:metric}
For each method, a cooperative policy was learned five times with different random seeds, and we evaluated each learned policy on a set of 100 new episodes. The following metrics were calculated and averaged over all episodes to measure the fairness, efficiency, and safety of multi-robot navigation by each method.
\begin{itemize}
   \item \textbf{Success Rate (SR)} measures the proportion of successful evaluation episodes, which indicates a method's safety.
   \item \textbf{Makespan (MS)} is the length of a valid trajectory $l^\pi$ for successful episodes, which indicates a method's efficiency.
   \item \textbf{Variance of Delays (VD)} is our fairness indicator that was introduced in Sec.~\ref{sec:preliminaries}. It is calculated for each successful episode as the variance of the differences between the lengths of a cooperative trajectory $l^\pi_i$ and the best possible trajectory $\tau^{\mu}_i$ obtained by the solitary policy.
   \item \textbf{Maximum Delay (MAXD) and Mean Delay (MEAND)} are the maximum and mean delay for each successful episode, thus giving another set of efficiency metrics.
\end{itemize}
\subsection{Results}
\begin{table*}[t]
    \centering
    \caption{Quantitative results. SR: success rate ($\uparrow$); MS: makespan ($\downarrow$); VD: variance of delays ($\downarrow$); MAXD: maximum delay ($\downarrow$); MEAND: mean delay ($\downarrow$). Each SR score is the average over five trials for 100 episodes with different random seeds, while the other scores are the averages over the successful episodes among them.}
\scalebox{0.9}{
    \begin{tabular}{@{}lccccccccccccccc@{}}
    \toprule 
    & \multicolumn{5}{c}{\textsc{Uniform-8-25}} 
    & \multicolumn{5}{c}{\textsc{Uniform-12-25}} &
    \multicolumn{5}{c}{\textsc{Uniform-16-25}} \\
    \cmidrule(lr){2-6}  \cmidrule(lr){7-11}  \cmidrule(lr){12-16} 
    & SR & MS & VD & MAXD & MEAND & SR & MS & VD & MAXD & MEAND & SR & MS & VD & MAXD & MEAND \\
    \midrule
        FEN~\citep{jiang2019learning}& 95.4 & \textbf{22.2} & 10.5 & 5.11 & 1.26 & 37.0 & 27.2 & 15.8 & 8.77 & 2.46 & 17.0 & 32.0 & 25.5 & 13.1 & 3.42 \\ 
        SOTO~\citep{zimmer2021learning}& 17.4 & 50.5 & 185 & 33 & 9.36 & 0.80 & 44.2 & 78.5 & 27.8 & 9.31 & 0 & N/A & N/A & N/A & N/A\\ 
        Nav-Only & \textbf{98.6} & 22.3 & 13.3 & 4.85 & 1.26 & \textbf{93.0} & \textbf{26.1} & 13.1 & 7.52 & \textbf{1.74} & 75.8 & \textbf{30.9} & \textbf{20.9} & \textbf{11.5} & \textbf{2.86}\\ 
        \midrule
        \textbf{NCF2 (Ours)} & 98.0 & \textbf{22.2} & \textbf{5.63} & \textbf{4.51} & \textbf{1.25} & 92.6 & 26.7 & \textbf{10.0} & \textbf{7.29} & 1.89 & \textbf{77.8} & 32.1 & 22.0 & 12.3 & 3.02\\ 
        \bottomrule
    \toprule 
    & \multicolumn{5}{c}{\textsc{Uniform-8-50}} 
    & \multicolumn{5}{c}{\textsc{Uniform-12-50}} &
    \multicolumn{5}{c}{\textsc{Uniform-16-50}} \\
    \cmidrule(lr){2-6}  \cmidrule(lr){7-11}  \cmidrule(lr){12-16} 
    & SR & MS & VD & MAXD & MEAND & SR & MS & VD & MAXD & MEAND & SR & MS & VD & MAXD & MEAND \\
    \midrule
        FEN~\citep{jiang2019learning}& 63.4 & 31.6 & 28.9 & 11.3 & 3.18 & 26.4 & 40.1 & 68.4 & 18.8 & 4.79 & 0 & N/A & N/A & N/A & N/A\\ 
        SOTO~\citep{zimmer2021learning}& 3.20 & 78.1 & 500 & 60.5 & 23.1 & 0 & N/A & N/A & N/A & N/A & 0 & N/A & N/A & N/A & N/A\\ 
        Nav-Only & 89.4 & \textbf{29.6} & \textbf{21.1} & \textbf{8.64} & \textbf{2.40} & 65.4 & \textbf{38.0} & 54.6 & \textbf{16.9} & 4.37 & 30.8 & 50.1 & 101 & 28.1 & \textbf{6.39}\\ 
        \midrule
        \textbf{NCF2 (Ours)} & \textbf{91.7} & 30.3 & 24.1 & 9.34 & 2.64 & \textbf{71.4} & 38.2 & \textbf{47.0} & 17.2 & \textbf{4.36} & \textbf{40.8} & \textbf{48.3} & \textbf{85.8} & \textbf{26.1} & 6.55\\ 
        \bottomrule
    \toprule 
    & \multicolumn{5}{c}{\textsc{Corner-8-25}} 
    & \multicolumn{5}{c}{\textsc{Corner-12-25}} &
    \multicolumn{5}{c}{\textsc{Corner-16-25}} \\
    \cmidrule(lr){2-6}  \cmidrule(lr){7-11}  \cmidrule(lr){12-16} 
    & SR & MS & VD & MAXD & MEAND & SR & MS & VD & MAXD & MEAND & SR & MS & VD & MAXD & MEAND \\
    \midrule
        FEN~\citep{jiang2019learning}& 73.8 & \textbf{34.8} & \textbf{37.8} & \textbf{14.9} & \textbf{5.84} & 29.6 & \textbf{48.4} & \textbf{73.9} & \textbf{28.0} & \textbf{15.6} & 8.80 & 77.6 & 283 & 57.7 & 35.3\\ 
        SOTO~\citep{zimmer2021learning}& 17.2 & 63 & 140 & 42.8 & 24.4 & 0 & N/A & N/A & N/A & N/A & 0 & N/A & N/A & N/A & N/A\\ 
        Nav-Only & 91.4 & 42.2 & 85.5 & 23.3 & 14.5 & 77.6 & 58.9 & 133.9 & 39.6 & 25.2 & \textbf{52.0} & 64.4 & 170 & 44.3 & 24.5\\ 
        \midrule
        \textbf{NCF2 (Ours)} & \textbf{93.8} & 43.7 & 85.0 & 24.8 & 15.7 & \textbf{86.4} & 50.7 & 74.6 & 31.0 & 19.5 & 49.6 & \textbf{55.5} & \textbf{101} & \textbf{35.1} & \textbf{18.5}\\ 
        \bottomrule
    \toprule 
    & \multicolumn{5}{c}{\textsc{Corner-8-50}} 
    & \multicolumn{5}{c}{\textsc{Corner-12-50}} &
    \multicolumn{5}{c}{\textsc{Corner-16-50}} \\
    \cmidrule(lr){2-6}  \cmidrule(lr){7-11}  \cmidrule(lr){12-16} 
    & SR & MS & VD & MAXD & MEAND & SR & MS & VD & MAXD & MEAND & SR & MS & VD & MAXD & MEAND \\
    \midrule
        FEN~\citep{jiang2019learning}& 44.4 & \textbf{45.3} & \textbf{64.2} & \textbf{21.8} & \textbf{10.2} & 8.00 & 79.9 & 359 & 59.2 & \textbf{30.7} & 0 & N/A & N/A & N/A & N/A\\ 
        SOTO~\citep{zimmer2021learning}& 0 & N/A & N/A & N/A & N/A & 0 & N/A & N/A & N/A & N/A & 0 & N/A & N/A & N/A & N/A\\ 
        Nav-Only & 54.4 & 68.0 & 285 & 46.8 & 29.4 & 20.8 & 78.2 & 323 & 57.1 & 33.2 & 4.00 & 82.6 & 325 & 61.3 & 33.0\\ 
        \midrule
        \textbf{NCF2 (Ours)} &\textbf{70.6} & 66.7 & 271 & 45.5 & 28.1 & \textbf{40.4} & \textbf{76.3} & \textbf{288} & \textbf{54.6} & 32.6 & \textbf{9.20} & \textbf{79.1} & \textbf{273} & \textbf{56.6} & \textbf{31.1} \\ 
        \bottomrule
    \end{tabular}
    }
    \label{tab:results}
\end{table*}

\paragraph{Quantitative Results} Tab.~\ref{tab:results} summarizes the quantitative comparison results. Overall, the proposed \methodabb performed well on every metric. In contrast, for challenging environments with 50 obstacles and more than 12 agents, it was hard or almost impossible for the baseline methods to successfully complete the navigation tasks. For SOTO, this was possibly because it is difficult to resolve a credit assignment problem by learning a team-oriented sub-policy just to balance the cumulative rewards among neighboring agents. FEN could perform better than SOTO and was sometimes comparable to or better than \methodabb in terms of the MS, VD, MAXD, and MEAND scores. Nevertheless, its SR scores were quite limited as the numbers of agents and obstacles both increased; that is, fair-delay navigation was possible only in a limited number of successful episodes. For FEN, we also observed many situations in which only the efficiency-oriented sub-policy was selected by the controller, rather than other sub-policies that would have allowed agents to give way to each other, thus limiting the SR. Nav-Only, the degraded version of \methodabb, demonstrated comparable success rates in environments that were not very difficult (\eg, \textsc{Uniform-8-25, Uniform-12-25, Uniform-16-25, Corner-8-25, Corner-16-25}). Nevertheless, its VD scores were worse than those of \methodabb, especially in \textsc{Corner} environments, because of the lack of a mechanism to maintain fairness regarding delays. Interestingly, in difficult environments with 50 obstacles, \methodabb could succeed in more episodes than Nav-Only, apart from the fairness. 
This was because the \methodabb policy gave agents a wider choice of actions, including ``staying still,'' and allowed the agents to consider the actions of selected agents who will move in the next step. On the other hand, the Nav-Only agents had to consider the actions of all neighboring agents. This limited the options for the Nav-Only agents' next actions and could result in collisions or timeouts in congested environments.

\paragraph{Qualitative Results} Fig.~\ref{fig:results} shows some visual examples of solutions by each method. For \methodabb, agents could stop properly when another agent was moving just ahead of them, as indicated by gray circles on their trajectories. On the other hand, agents tended to get stuck (gray triangles) or even collided with other obstacles (gray crosses) with the other baselines.

\begin{table}[t]
\centering
\caption{Ablation study. (i) w/o improvements: $\xi^j_t=1$ to disable counterfactual reasoning. (ii) w/o limited comm.: $\mathcal{N}^i_t=\{1,\ldots,N\}$ to enforce agents to exchange messages with all other agents. (iii) fixed priority: $\rho^i_t=i-1$ to avoid using patience scores.}
  \begin{tabular}{lccccc}
  \toprule
  & SR & MS & VD & MAXD & MEAND \\
  \midrule
  (i) w/o improvements & 84.0 & 58.0 & 99.5 & 36.8 & 20.4\\
  (ii) w/o limited comm. & 79.2 & 55.3 & 113 & 34.9 & 19.6\\
  (iii) fixed priority & 62.8 & \textbf{50.3} & 86.4 & \textbf{30.9} & \textbf{16.4}\\
  \midrule
  \textbf{NCF2} &\textbf{86.4} & 50.7 & \textbf{74.6} & 31.0 & 19.5  \\
  \bottomrule
  \end{tabular}
  \label{tab:abb}
\end{table}

\paragraph{Ablation Study} We additionally evaluated the following degraded versions of \methodabb in the \textsc{Corner-12-25} environment to demonstrate the importance of each component: (i) \textbf{w/o improvements}, which used $\xi^j_t = 1$ for all $j$ and $t$ in Eq.~(\ref{eq:fair-efficient-reward}) to disable the counterfactual reasoning; (ii) \textbf{w/o limited communications}, which used $\mathcal{N}^i_t=\{1,\ldots,N\}$ for all $i$ and $t$ to examine how allowing message exchanges beyond an agent's proximity affected the performance; and (iii) \textbf{fixed priority}, to examine how the performance changed if the patience was given deterministically as $\rho^i_t=i-1$ rather than by considering an agent's actual patience. As listed in Tab.~\ref{tab:abb}, each of these versions showed decreased performance as compared to the proposed method. While the MEAND (mean of delays) did not differ among the methods, the VD (variance of delays) showed a considerable gap. This result indicates that the counterfactual reasoning, limited communication, and use of patience scores for improved fairness regarding delays were all important.

\paragraph{Failure Cases and Limitations}
\looseness=-1
Although \methodabb worked well in many cases, our work has several limitations. We found that the \methodabb policy sometimes failed to resolve situations when agents were crowded into a small area. This was because the actions to avoid collisions are inherently limited in highly congested environments, thus making it difficult to expand other agents' action choices simply by taking cooperative actions and improving the fairness. A possible way to mitigate such problems in congested environments would be to incorporate global planning that searches the environment for low-cost, feasible paths. This approach can address the myopia of local planning and make it possible to actively bypass congested areas. Finally, while we assumed the state transitions and agent observations to be deterministic, they are stochastic in practice because of various noise sources~\cite{thrun2002probabilistic}. Extensions of our work to stochastic environments would be an interesting future direction.
\section{Related Work}
Fairness has been actively studied in MARL. Typical tasks include job scheduling, the Matthew effect, traffic control, and so on in a cooperative setting~\cite{jiang2019learning,umer2020learning,zimmer2021learning}, and fair resource allocation~\cite{zhang2014fairness,elzayn2019fair,d2020fairness} and social dilemmas~\cite{leibo2017multi,hughes2018inequity,jaques2019social} in a mixed setting. More recently, the application of prediction-based fairness~\cite{corbett2018measure,pmlr-v81-buolamwini18a} was studied in a multi-agent setting  \cite{grupen2022cooperative}. Two prior studies that provided baselines in our experiments~\cite{jiang2019learning,zimmer2021learning} especially addressed the tradeoff between fairness and efficiency. Nevertheless, these methods have limitations in resolving complex credit assignment problems to improve fairness while maintaining navigation safety and efficiency, as demonstrated in the experiments.

\looseness=-1
On the other hand, the notion of counterfactuals has recently attracted much attention as a powerful tool to address complex multi-agent credit assignment problems. Since a pioneering work~\cite{foerster2018counterfactual} that used counterfactuals to compute the baseline for the actor-critic algorithm, more recent works have extended counterfactuals with a graph convolution communication~\cite{su2020counterfactual}, Shapley counterfactual credits~\cite{li2021shapley}, and simulation reasoning for fully-decentralized counterfactual MARL~\cite{yuan2022counterfactual}. However, all these methods have been found effective only in discrete action spaces. For some methods, the specification of default actions is nontrivial in a continuous space~\cite{li2021shapley,yuan2022counterfactual}. Moreover, all but the last one~\cite{yuan2022counterfactual} require a centralized critic and have limited scalability with respect to the number of agents. 

Finally, multi-robot navigation is a popular task in the robotics community. It has been addressed analytically~\cite{van2008reciprocal,alonso2013optimal,bareiss2015generalized}, via MAPF~\cite{honig2018trajectory,li2019multi,luis2020online}, or via machine learning including MARL~\cite{godoy2016implicit,long2018towards,fan2020distributed,xu2021human}, to name a few approaches. In any case, the main challenges in these studies have been successful collision avoidance for safety and improvement in the trade-off between safety and efficiency, without considering how to achieve fairness as well.
\section{Conclusion}
We have introduced \problem, a novel extension of fairness-aware MARL, with the objective of finding a set of trajectories for a team of agents. The trajectories should be collision-free, short, and as fair as possible in terms of temporal delays as compared to the respective shortest possible trajectories when ignoring the presence of other agents. To address this problem, we proposed a new navigation algorithm, called \methodabb, that enables counterfactual inference on whether each agent should move or stay still to improve the fairness regarding delays while maintaining travel efficiency and safety. In terms of those criteria, the proposed method outperformed state-of-the-art fairness-aware MARL methods in challenging multi-robot navigation environments.

\balance
\bibliographystyle{ACM-Reference-Format} 
\bibliography{aamas2023_conference}

\newpage
\appendix
\begin{figure*}[t]
    \begin{minipage}[b]{0.49\linewidth}
        \centering
        \includegraphics[width=1.0\linewidth]{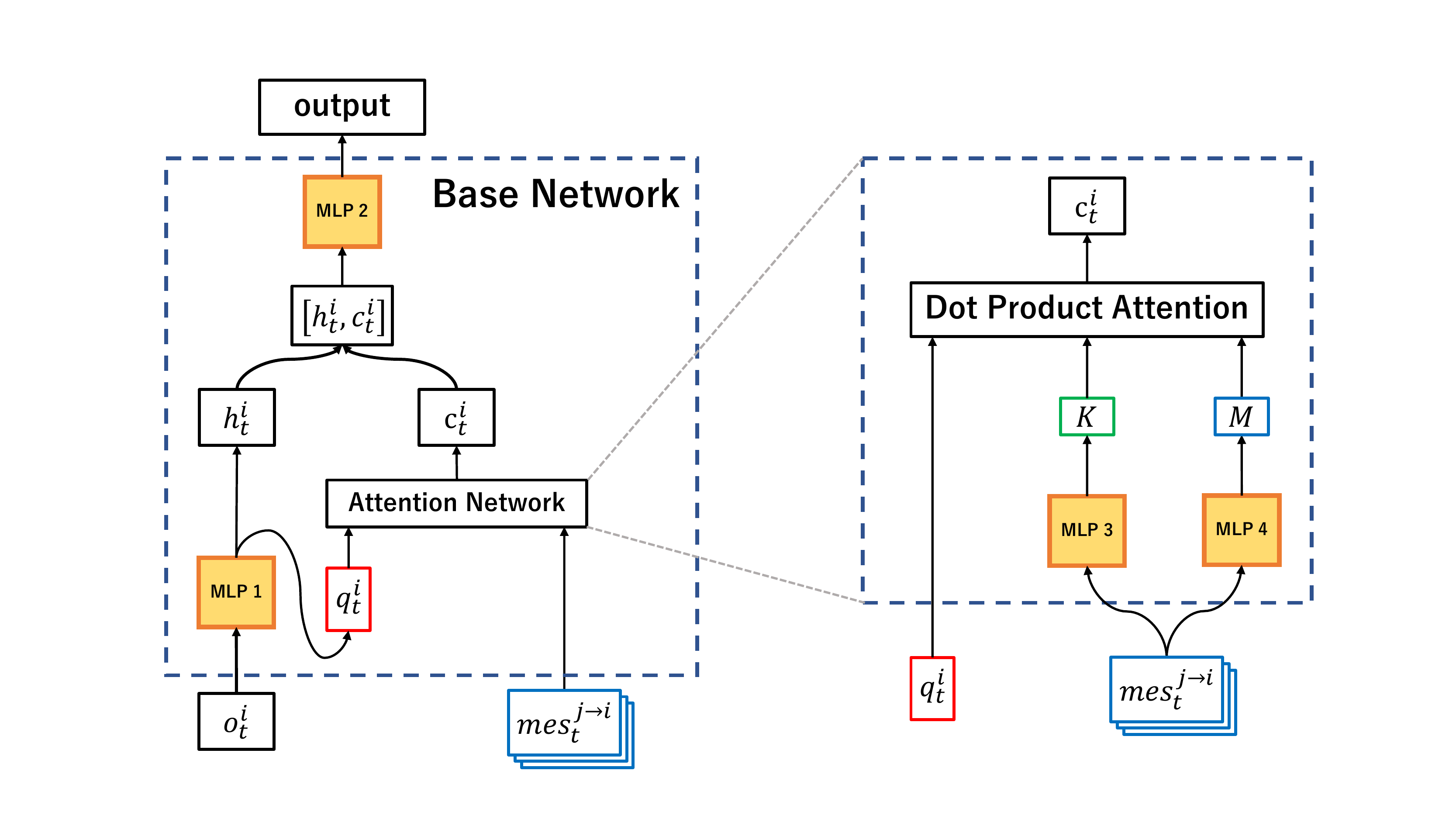}
    \end{minipage}
    \begin{minipage}[b]{0.49\linewidth}
        \centering
        \includegraphics[width=1.0\linewidth]{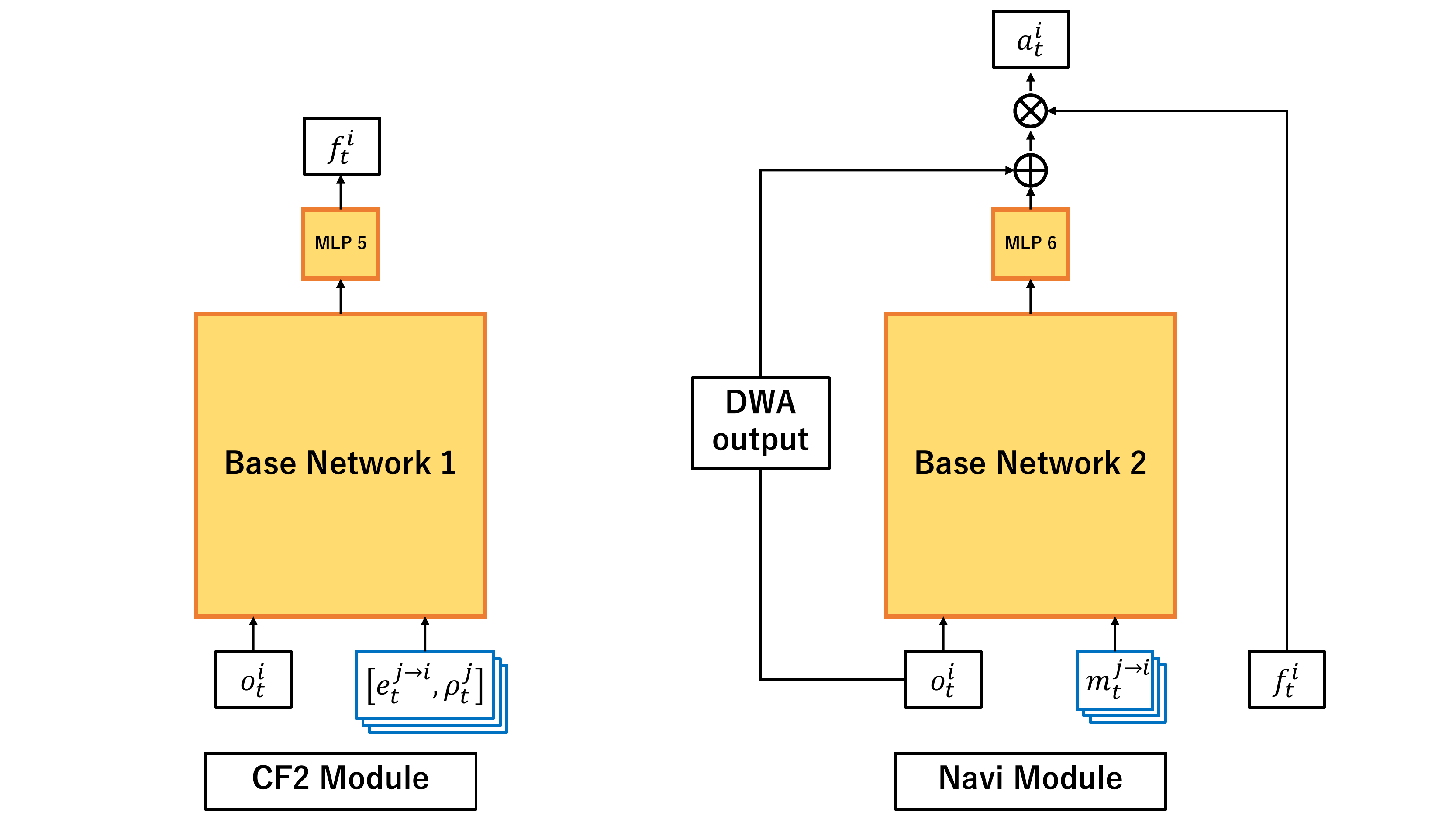}
    \end{minipage}
    \caption{Network architecture for \methodabb}
    \label{fig:base_net}
\end{figure*}

\section{Reward Analysis}
\label{sec:reward}
As described in Sec.~2, our original fairness objective was to make the variance of delays, $\frac{1}{N}\sum_i(\delta_i - \frac{1}{N}\sum_j \delta_j)^2$, as low as possible. In Sec.~3, we introduced the patience $\rho^i_t$ for each timestep $t$ as a proxy for the delay $\delta_i$. Here, we relate the fairness-efficiency reward $\tilde{r}^i_t$ in Eq.~(8) to the fair-delay objective. By letting $\bar{\rho}^j_t = \frac{\rho^j_t}{\sum_{k\in\mathcal{N}^i_j}\rho^k_t}$, the first summation term of $\tilde{r}^i_t$ can be transformed as follows:
\begin{eqnarray}
\frac{\sum_{j\in\mathcal{N}^i_t}(\rho^j_t - \rho^i_t)\cdot \xi^j_t}{\sum_{j\in\mathcal{N}^i_t} \rho^j_t}  &=& \sum_{j\in\mathcal{N}^i_t} (\bar{\rho}^j_t - \bar{\rho}^i_t) \cdot \xi^j_t\\ 
&=& \sum_{j\in\mathcal{N}^i_t}\xi^j_t\cdot\left(\frac{\sum_{j\in\mathcal{N}^i_t} \xi^j_t\bar{\rho}^j_t}{\sum_{j\in\mathcal{N}^i_t}\xi^j_t} - \bar{\rho}^i_t\right) \\
&=&\frac{\left(\sum_{j\in\mathcal{N}^i_t} \xi^j_t\right)^2}{2\xi^i_t}\cdot \frac{2\xi^i_t\left(\mu - \bar{\rho}^i_t\right)}{\sum_{j\in\mathcal{N}^i_t}\xi^j_t}\\
&=& -K\cdot \frac{\partial}{\partial \bar{\rho}^i_t}V.
\label{eq:derivative-of-variance}
\end{eqnarray}
Here, $\mu=\frac{\sum_{j\in\mathcal{N}^i_t}\xi^j_t\bar{\rho}^j_t}{\sum_{j\in\mathcal{N}^i_t}\xi^j_t}$, $K=\frac{\left(\sum_{j\in\mathcal{N}^i_t} \xi^j_t\right)^2}{2\xi^i_t}$, and $V=\frac{\sum_{j\in\mathcal{N}^i_t} \xi_j\left(\bar{\rho}^j_t - \mu\right)^2}{\sum_{j\in\mathcal{N}^i_t}\xi^j_t}$, where $V$ is the weighted variance for $\{\bar{\rho}^i_t\}$ with a weight given by $\{\xi^i_t\}$. This result can be interpreted that an agent will receive a higher reward if stopping will decrease the variance of the (normalized) patience for certain agents that are in its proximity, as specified by $\mathcal{N}^i_t$, and whose patience can improve. Note that the patience is suppressed by $\frac{1}{\xi^i_t}$ in $K$ so as not to ignore $P_i$'s own room for improvement. Also, the reward will be low when there is little room for overall improvement as indicated by $\sum_{j\in\mathcal{N}^i_t}\xi^j_t$ in $K$. If we assume that $\mathcal{N}^i_t=\{1,\ldots,N\}$ and $\forall j,\;\xi^j_t=1$, $V$ in Eq.~(\ref{eq:derivative-of-variance}) reduces to the variance of the original patience $\rho^i_t$ of all agents. Nevertheless, our ablation study results (see Tab. 2) empirically confirmed that limited communication via $\mathcal{N}^i_t$ and improvement via $\xi^i_t$ are both crucial for better performance.

\section{Complete Implementation Details}
\subsection{Message Encoding}
\label{sec:encoding}
We encode a collection of messages, $\{e^{j\rightarrow i}_t\mid j \in \mathcal{N}^i_t\}$ and $\{m^{j\rightarrow i}_t\mid j\in\mathcal{N}^i_t\}$, in a single fixed-length vector by using the scaled dot-product attention mechanism~\citep{vaswani2017attention}. Here, we descrbe the formulation for encoding $\{m^{j\rightarrow i}_t\mid j \in \mathcal{N}^i_t\}$, but exactly the same approach holds for $e^{j\rightarrow i}_t$. Let $M_\mathrm{c}, M_\mathrm{n}$ be the respective vertical stacks of elements $f^j_t\Delta(s^j_t, s^i_t)$ and $f^j_t\Delta(\hat{s}^j_{t+1}, s^i_t)$ in $\{m^{j\rightarrow i}_t\mid \mathcal{N}^i_t\}$, where $t$ is omitted hereafter for simplicity. These stacks are encoded separately with independent attention encoders and concatenated to form a fixed-length vector. By letting $Q_\mathrm{c}, K_\mathrm{c}, V_\mathrm{c}$ (resp. $Q_\mathrm{n}, K_\mathrm{n}, V_\mathrm{n}$) be the query, key, and value obtained from $M_\mathrm{c}$ (resp. $M_\mathrm{n}$), the encoded vector is given as follows:
\begin{equation}
   m = \left[\sigma\left(\frac{Q_\mathrm{c}K_\mathrm{c}^\top}{\sqrt{d}}\right)V_\mathrm{c};\sigma\left(\frac{Q_\mathrm{n}K_\mathrm{n}^\top}{\sqrt{d}}\right)V_\mathrm{n}\right],
\end{equation}
where $\sigma(\cdot)$ denotes the softmax function and $d$ is the key dimension.

\subsection{Network Architecture}
\label{sec:network}

Fig.~\ref{fig:base_net} shows the network architecture for \methodabb. Its specific parameters, as well as those for FEN and SOTO, are summarized in Tab.~\ref{tab:hp_arch}.

\begin{table}[h]
\centering
\caption{Hyper-parameters for the network architectures.}
\label{tab:hp_arch}
\begin{tabular}{@{}lc@{}}
\toprule 
Name & Value \\
\midrule 
Common Architecture & \\
\multicolumn{1}{r}{MLP 1} & FC(256)-ReLU\\
\multicolumn{1}{r}{MLP 2} & FC(256)-ReLU\\
\multicolumn{1}{r}{MLP 3} & FC(256)-ReLU-FC(24)\\
\multicolumn{1}{r}{MLP 4} & FC(256)-ReLU-FC(24)\\
\midrule 
NCF2 & \\
\multicolumn{1}{r}{MLP 5} & FC(2)-Softmax\\
\midrule
FEN & \\
\multicolumn{1}{r}{MLP 5} & FC(4)-Softmax\\
\multicolumn{1}{r}{Number of sub-policies} & 4 \\
\midrule
SOTO & \\
\multicolumn{1}{r}{MLP 6} & FC(2)\\
\bottomrule
\end{tabular}
\end{table}

\subsection{Distributed Learning Setup}
\begin{figure}
    \centering
    \includegraphics[width=.9\linewidth]{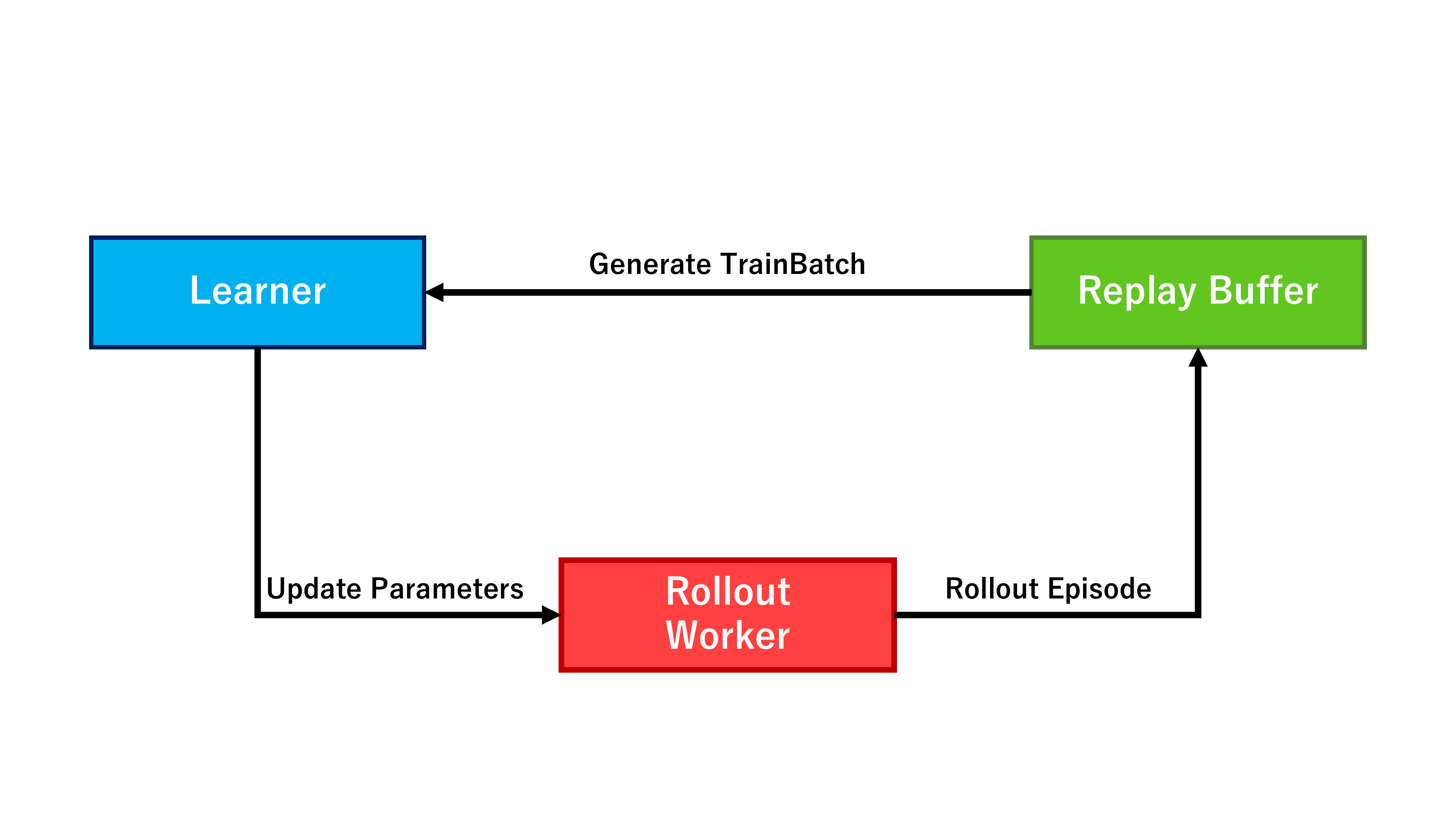}
    \caption{Diagram of our distributed learning setup}
    \label{fig:dist_architecture}
\end{figure}

\begin{table}[h]
\centering
\caption{Hyperparameters for the learning setup.}
\label{tab:hp_learning}
\begin{tabular}{@{}lc@{}}
\toprule 
Name & Value \\
\midrule 
Common Parameters & \\
\multicolumn{1}{r}{Discount factor}& 0.95 \\ 
\multicolumn{1}{r}{Initial temperature}& 0.01 \\
\multicolumn{1}{r}{Target network update rate}& 0.005 \\
\multicolumn{1}{r}{Target update interval}& 1 \\
\multicolumn{1}{r}{Learning rate}& 0.001 \\
\multicolumn{1}{r}{Batch size}& 256 \\
\multicolumn{1}{r}{Replay buffer size}& 1500000 \\
\multicolumn{1}{r}{Iterations}& 1000000 \\
\multicolumn{1}{r}{Iterations for critic warm-up}& 10000 \\
\midrule 
FEN & \\
\multicolumn{1}{r}{Interval for sub-policy selection} & 5 \\
\multicolumn{1}{r}{Learning rate for controller} & 0.0005 \\
\midrule
SOTO & \\
\multicolumn{1}{r}{GGF weight} & $\left(1, 0.5, 0.25, ... \right)$ \\
\bottomrule
\end{tabular}
\end{table}
As described in Sec. 4.2, our SAC implementation adopted the distributed experience-generating strategy of Ape-X~\cite{horgan2018distributed}. As shown in Fig.~\ref{fig:dist_architecture}, the implementation has a learner, a rollout worker, and a replay buffer. The rollout worker gathers a collection of experience with the current policy and sends it to the replay buffer, which then generates a training batch for the learner. Next, the learner uses the batch to update the network parameters and sends the parameters to the rollout worker. These procedures are performed completely in parallel. The hyperparameters are listed in Table~\ref{tab:hp_learning}. Note that we set a smaller learning rate for the FEN controller; this was necessary to allow it to learn meaningful choices based on sub-policies that had been learned ahead of time.

\end{document}